\documentstyle[aps,epsfig,multicol,fancyheadings]{revtex}
\begin{document}

\title{Extremal statistics in the energetics of domain walls}

\author{E.\ T.\ Sepp\"al\"a,$^1$  M.\ J.\ Alava,$^1$ and P.\ M.\ Duxbury$^2$}

\address{$^1$Helsinki University of Technology, Laboratory of Physics,
P.O.Box 1100, FIN-02015 HUT, Finland }
\address{ $^2$Dept. of Physics and Astronomy and Center for 
Fundamental Materials Research,\\
Michigan State Univ., E. Lansing, MI 48824-1116, U.S.A.}

\maketitle

\begin{abstract}
\noindent
We study at $T=0$ the minimum energy of a domain wall and its gap to
the first excited state concentrating on two-dimensional random-bond
Ising magnets. The average gap scales as $\Delta E_1 \sim L^\theta
f(N_z)$, where $f(y) \sim [\ln y]^{-1/2}$, $\theta$ is the energy
fluctuation exponent, $L$ length scale, and $N_z$ the number of energy
valleys.  The logarithmic scaling is due to extremal statistics, which
is illustrated by mapping the problem into the Kardar-Parisi-Zhang
roughening process.  It follows that the susceptibility of domain
walls has also a logarithmic dependence on system size.
\end{abstract}

\noindent {\it PACS \# \ \ 05.70.Np, 75.50.Lk, 68.35.Ct, 64.60.Ht}

\begin{multicols}{2}[]
\narrowtext

The energy landscapes of random systems are often assumed to be
described at low temperatures by scaling exponents that follow from
the behavior of the ground states.  In the renormalization group (RG)
language this means that temperature is an irrelevant variable. In
most quenched random systems, the energy landscape contains many
low-lying metastable minima separated by high barriers. Examples can
be found in the realm of random magnets, the most famous one being of
course spin glasses~\cite{review}.  The dynamical behavior at finite
temperatures, as a result of a temperature change or the application
of an external field, will naturally depend on the associated barriers
and energy differences between the minima.

It is often assumed that energy differences or barriers between
configurations ($\delta E$) relate to the length $l$ involved by a
scaling relation $\delta E \sim l^\theta$, where $\theta$ is an energy
fluctuation exponent. It measures the dependence of the first
non-analytic correction to the ground state or free energy on the
length scale. Here we show that for extended manifolds, or Ising magnet
domain walls (DW) [equivalent to directed polymers (DP) in
(1+1)-dimensions] the energy difference between the ground state energy
and the next state (``first excited state'') follows from {\it
extremal statistics}. This is due to the fact that, usually, one can
assume that the energy landscape, at large enough scales, consists of
many {\it independent} valleys.  Finding the gap between the minimum
and the second-most favorable state is then a straightforward extremal
statistics problem as is the simpler one of the minimum of all the
independent valley energies.  The extreme statistics leads to
logarithmic factors in the gap and minimum energies, which we show
also by numerical calculations. The same result can be applied also to
other disordered systems, where the energy landscape of DW can be
reduced to a one-dimensional form.  We also interpret the results in
the language of kinetic roughening, since DP maps into the
Kardar-Parisi-Zhang (KPZ) equation of growth
\cite{KPZ,Hah95,Barabasi}.  Finally, as an application we show that
the extremal statistics scaling shows up in the {\it susceptibility}
of DW.

Here we consider elastic manifolds at $T=0$ with quenched short-range,
e.g. point-like defects, randomness and in $d=(D+n), n=1$ dimensions,
in which $D$ is the dimension of the manifolds and $d$ is the
dimension of their embedding space.  The continuum Hamiltonian for
such an elastic manifold is
\begin{equation}
{\mathcal H} = \int \, \left[ \frac{\Gamma}{2} \{ \nabla z({\bf x})
\}^2 + V_r ({\bf x},z)\right] {\rm d}^D{\bf x},
\label{H}
\end{equation} 
where $z({\bf x})$ is the height of the interface and ${\bf x}$ is the
$D$ dimensional internal coordinate of the manifold. The first term in
the integrand is the elastic contribution with the corresponding
surface stiffness $\Gamma$ of the interface and the second term comes
from the random potential. For random manifolds we use quenched random
bond (RB) disorder, which means that the random potential is
delta-point correlated, i.e., $\langle V_r ({\bf x},z)V_r ({\bf
x'},z') \rangle =2 {\mathcal D} \delta ({\bf x}- {\bf x'})\delta
(z-z)$.  The geometric behavior of the manifold is characterized by
$w^2 = \left \langle \left[z({\bf x}) - \overline{z({\bf x})} \right
]^2 \right \rangle \sim L^{2 \zeta}$, where $L$ is the linear size of
the system and $\zeta$ is the corresponding roughness exponent. At low
temperatures in $(1+1)$ dimensions, due to the equivalence of DP in
random media~\cite{KPZ,Hah95} to the KPZ equation, the exact roughness
exponent reads $\zeta=2/3$~\cite{KPZ,Hah95,Barabasi}.  In higher
dimensions functional RG gives the approximate expression $\zeta
\simeq 0.208(4-D)$~\cite{Fis86} for RB DW.  Since the width of a
manifold grows as $L^\zeta$ it is expected that the number of
independent valleys~\cite{Mez,Hwa} is proportional to $N_z \sim
L_z/L^\zeta$.  At $T=0$ the total average minimum energy $\langle E_0
\rangle$ of an elastic manifold equals its free energy and grows
linearly with the manifold area $L^D$ and its fluctuations scale as
$\Delta E = \left \langle ( E_0- \langle E_0 \rangle )^2 \right
\rangle^{1/2} \sim L^\theta$, where $\theta = 2 \zeta +D
-2$~\cite{HuHe}.

Let us now derive analytically the scaling of the ``extreme
statistics'' contributions to the lowest minimum, $E_0$, and the gap
between two lowest minima, $\Delta E_1 = E_1 -E_0$.  We consider the
case of many independent valleys in the landscape $N_z > 1$, which
means that the DP can have an arbitrary starting or end point, and
that $L_z > L^\zeta$.  For the ``single valley'' boundary condition
case (one end of the manifold fixed), it is known numerically that
near its mean the distribution is Gaussian~\cite{KBM}. Hence we draw
the energies $E$ from the distribution
\begin{equation}
P(E) =  k \exp\left \{- \left({|E - \langle E \rangle | \over \Delta E} 
\right)^{\eta} \right \},
\label{distr}
\end{equation}
where $\langle E \rangle \sim L^D$ is the average energy of the
manifold and $\Delta E \sim L^\theta$ measures its fluctuations and
$k$ normalizes the integral so $k\sim 1/L^{\theta}$.  The exponent
$\eta$ is not constant~\cite{KBM,Hah95}.  Near the peak $\eta=2$.  In
the low energy tail numerical simulations indicate that $\eta_-
\approx 1.6$, while in the high energy tail the best estimate is
$\eta_+ \approx 2.4$~\cite{KBM}.  At this stage we allow $\eta$ to be
variable but note that it is the behavior near the mean and the {\it
low energy tail} which is the most important in this calculation.  In
a system with $N_z \sim L_z/L^\zeta$ independent local minima the
probability that the global minimum has energy $E$ is given by,
\begin{equation}
L_{N_z}(E) = N_z P(E) \left \{1-C_1(E) \right \}^{N_z-1},
\label{global}
\end{equation}
where $C_1(E) = \int_{-\infty}^E P(\epsilon) \,
d\epsilon$~\cite{Galambos}.  The gap $\Delta E_1$ follows similarly.
Its distribution, $G_{N_z}(\Delta E_1,E)$ is given by
\begin{eqnarray}
G_{N_z}(\Delta E_1,E) = N_z(N_z-1) P(E) P(E+\Delta E_1) \nonumber\\
\{1-C_1(E+\Delta E_1)\}^{N_z-2}.
\label{gapdis}
\end{eqnarray}
$G_{N_z}(\Delta E_1,E)$ is the probability that if the lowest energy
manifold has an energy $E$, then the gap to the next lowest energy
level is $\Delta E_1$.  The average value of the global minimum is
given by
\begin{equation}
\langle E_0 \rangle = \int_{-\infty}^{\infty} E L_{N_z}(E) \,dE,
\end{equation}
which is not analytically integrable.  The typical value of the lowest
energy may be estimated using an {\it extreme scaling} estimate. It
follows from the fact the term the inside $\{ \, \}$ in (\ref{global})
becomes unity if $C_1$ is small enough. It has proven useful in other
contexts, for example breakdown of random networks, and reads
here~\cite{DBL},
\begin{equation}
1/k N_z P(\langle E_0 \rangle) \approx 1
\label{estimate}
\end{equation}
which yields,
\begin{equation}
\langle E_0 \rangle \approx \langle E \rangle - \Delta E 
\{ \ln(N_z) \}^{1/\eta}, 
\label{typicalene}
\end{equation}
where $\Delta E \sim L^\theta$. 

To estimate the typical value of the gap, we use similarly to
(\ref{estimate})
\begin{equation}
1/k^2 {N_z(N_z-1)} P(\langle E_0 \rangle ) P(\langle E_0\rangle +
\langle \Delta E_1 \rangle ) \approx 1,
\end{equation}
which with (\ref{typicalene}) and the fact that $| \langle \Delta E_1
\rangle | \ll | \langle E_0 \rangle| $ yields,
\begin{equation}
\langle \Delta E_1 \rangle \approx {  \Delta E^{\eta} \over 
\eta (\langle E \rangle - \langle E_0\rangle)^{\eta-1}} 
\approx {  \Delta E \over 
\eta \{\ln(N_z)\}^{(\eta-1)/\eta}}.
\label{gap}
\end{equation}
We thus find that in addition to the usual sample to sample variations
in the energy ($\Delta E \sim L^{\theta}$) there is a slow reduction
in the gap which scales as $\{\ln (N_z)\}^{-(\eta-1)/\eta}$, provided
$N_z > 1$.  Our case is closely related to the {\it weakly broken}
replica symmetry~\cite{Parisi} of DP, see also
Ref.~\cite{boumez}, where the relation between replica methods and
extremal statistics is discussed.

\begin{figure}[f]
\centerline{\epsfig{file=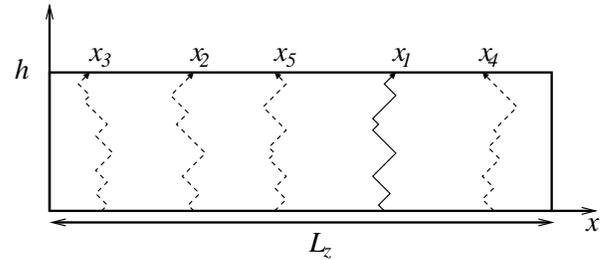,width=8cm}}
\caption{The relation between DP and growing interfaces.  KPZ
interface is growing so that $h$ increases and DPs in independent
valleys equal the n-th fastest arrival times of the interface to a
prefixed height $h$, at $x_n$, at times $t(x_n)$ in a system with
width $L_z$.  The solid line describes the fastest polymer, which ends
at $x_1$. The dashed lines describe the next fastest polymers.}
\label{fig1}
\end{figure}

The $(1+1)$ dimensional DW maps, in the continuum limit, to the KPZ
equation by associating the minimum energy of a DW with the minimum
{\it arrival time} $t_1 \equiv E_0$ of a KPZ-surface to height
$h$. The connection is illustrated in Fig.~\ref{fig1} in the limit of
many valleys $N_z > 1$.  The minimal path of the DW with the endpoint
$z(L)$ equals the path by which the interface reaches $h=L$, at location
$x_1 =z$ and at time $t_1=E_0$. Thus $t_1$ attains a logarithmic correction,
from Eq.~(\ref{typicalene}), of size $-h^\beta \{\ln(L_z/h^{1/z})
\}^{1/\eta}$, where $\beta = 1/3$ and $z=3/2$ are now the roughening
exponent and dynamical exponent of the KPZ universality class
\cite{KPZ}.  Consider now the second smallest arrival time $t_2$. In
the KPZ language of DP, if the path $x_2 (t')$ that gives $t_2$ is completely
independent of the $x_1 (t')$ that results in $t_1$ then $t_2$
and $x_2$ are related to a separate, independent valley of the
DP landscape.  The {\it
difference} $\Delta t = t_2 - t_1$ then equals $\Delta E_1$ of DW, and
likewise obeys extremal statistics, so that $\Delta t \sim h^\beta
[\ln (L_z/h^{1/z})]^{-(\eta-1)/\eta}$.  For growing surfaces this
limit is the {\it early stages} of growth, in which the correlation
length $\xi \ll L_z$, and therefore the arrival times, or DW energies,
are independent.

In order to check the scaling behavior of the gap energy (\ref{gap}),
we have done extensive exact ground state calculations of elastic
manifold in the two dimensional (2D) spin-half RB Ising
model, i.e. we take a nearest neighbor Ising model with random but
ferromagnetic couplings $J_{ij}>0$.  Calculations are performed by
varying both the parallel length $L$ and the height $L_z$ of systems
oriented in the \{10\} direction. The DW is imposed by antiperiodic
boundary conditions in the $z$-direction at $z=0$ and $z=L_z$.  The
elastic manifold is the interface, which divides the system in two
parts one containing up spins and the other containing down spins. At
$T=0$ the problem of finding the ground state DW is a global
optimization problem which is solved exactly using a mapping to the
minimum-cut maximum-flow problem.  The so-called push-and-relabel
method solves this problem efficiently and has been extensively
discussed elsewhere~\cite{Goltar88,Alavaetal,Seppala00}.

\begin{figure}[f]
\centerline{\epsfig{file=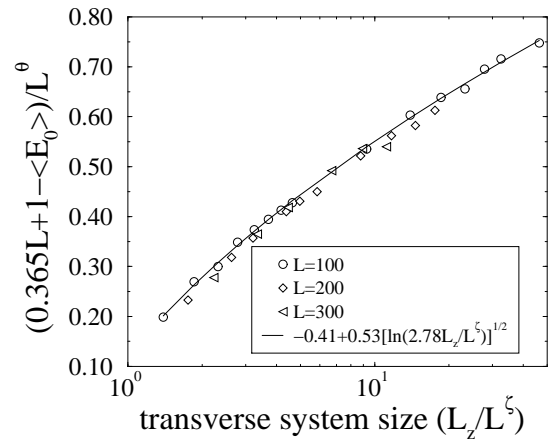,width=6cm,angle=-90}}
\caption{The scaling of the ground state energy $E_0$ as a function of
scaled transverse system size $L_z/L^{\zeta}$ for the system sizes
$L=$100, 200, and $300$. The line $-0.41+0.53 [ \ln (2.78
L_z/L^\zeta)]^{1/2}$ is a guide to the eye. We have subtracted the
expected dependence of $\langle E \rangle$ from $\langle E_0 \rangle$
(see the text).  We use in Figs.~\ref{fig2} to \ref{fig4} RB
disorder with $J_{ij,z} \in [0\--1]$ uniform distribution and
$J_{ij,x} = 0.5$. The number of realizations ranges from $N=500$ for
$L=300$, $L_z=500$ to $N=2000$ for $L=200$, $L_z=600$.}
\label{fig2}
\end{figure}

In order to control the average number of the minima $\langle N_z
\rangle \sim L_z/L^\zeta$ in a chosen system size, we set the initial
position of the interface $\overline{z}_0$ in a fixed size window at
height $\overline{z}_0/L_z \simeq \rm{const}$. If the ground state
interface is originally outside the window, with room only for a
single valley, it is neglected and a new configuration is created.
After the original ground state is found, with its energy $E_0$, the
lattice is reduced so that bonds in and above the window are neglected
and the new ground state, its $E_1$ and the corresponding gap energy
$\Delta E_1$, is found.  We studied at least $N=500$ realizations of
system sizes up to $L=300$ and $L_z=500$.  Fig.~\ref{fig2} starts the
discussion of the numerical data by showing how the ground state energy
$\langle E_0 \rangle$ behaves as a function of $L$ and $L_z$. The
scaling result (\ref{typicalene}) expects that the correction to the
energy follows a logarithmic dependence on $N_z$, which is confirmed
in the figure.  Note that the extraction of this correction from the data
requires an educated guess of how $\langle E \rangle$, the single
valley energy, behaves with $L$. We have used an Ansatz $\langle E
\rangle \sim aL +b$ with the values of $a$ and $b$ demonstrated in
Fig.~\ref{fig2}, so that the exponent value $\eta = 2$ corresponds to
a Gaussian distribution.  Due to the nature of the procedure it would
probably be possible to obtain a reasonable fit for e.g. $\eta =
\eta_-$ as well.

\begin{figure}[f]
\centerline{\epsfig{file=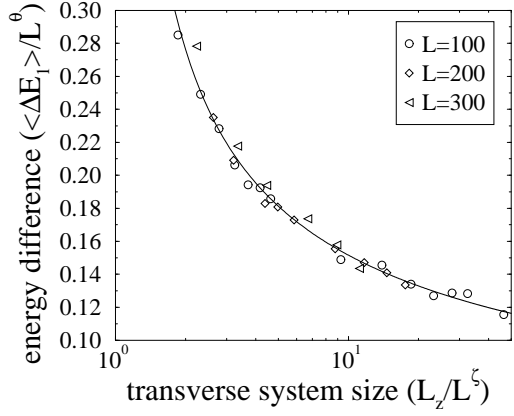,width=6cm,angle=-90}}
\caption{The scaling function $f(y)$ of the scaled disorder-average of
the energy difference $\langle \Delta E_1 \rangle /L^{\theta}$ as a
function of scaled transverse system size $L_z/L^{\zeta}$ for the
system sizes $L=$100, 200, and $300$, each with $\bar{z}_0/L_z \simeq
\rm{const}$. $\theta =1/3$, $\zeta =2/3$. The line has a shape $f(y) =
0.23 \ln(y)^{-1/2}$. The configurations are the same as in
Fig.~\ref{fig2}.  }
\label{fig3}
\end{figure}

For small sample sizes, $L_z < L^\zeta$ the value of the energy $E_0$
is affected by confinement.  Similarly, the gap is controlled by
confinement effects in this limit.  When $L_z$ is large there are many
independent valleys and extreme statistics effects are important,
hence we expect,
\begin{equation}
\langle \Delta E_1(L,L_z) \rangle \sim \left\{ \begin{array} {lll}
\tilde {f}(L_z),&\mbox{\hspace{5mm}}&L_z \ll L, \\
{L^{\theta}/ [\ln(L_z/L^{\zeta})]^{(\eta-1)/\eta}},& &L_z \gg L 
\end{array} \right.
\label{limits} 
\end{equation}
where we have used Eq.~(\ref{gap}) and $N_z \sim L_z/L^\zeta$.  
We attempt to collapse the data by using the reduced variables
$\langle \Delta E_1(L,L_z) \rangle/L^\theta$ versus $L_z/L^{\zeta}$
for various $L$ and $L_z$.  As seen in Fig.~\ref{fig3} we find a nice
agreement with the extreme scaling form, with the ratio $(\eta-1)/\eta
= 1/2$, i.e. by using a Gaussian distribution.

Next we consider the relation of the extremal statistics to the
susceptibility of these manifolds.  In the $D$-dimensional case
the susceptibility is defined by,
\begin{equation}
\chi = \lim_{h \to 0+} \left \langle \frac{\partial m}{\partial h} 
\right \rangle,
\label{Eqsuskis}
\end{equation} 
where the change in the magnetization of the whole $d$ dimensional
system is calculated in the limit of the vanishing external field from
the positive side~\cite{Seppala00,unpub} and the brackets imply a
disorder-average. We have recently shown that the general behavior
follows from a level-crossing phenomenon, which involves an extra
potential $V_h(z) = hz$, dependent on the height of the interface, in
the Hamiltonian~(\ref{H}), and $h$ is an applied external field to the
manifold. In any particular configuration when $h$ is varied the
manifold position changes in macroscopic 'jumps' \cite{Seppala00}, the
first one occurring at $h_1$.

One may write the susceptibility, Eq.~(\ref{Eqsuskis}), with the help
of the probability distribution of the fields $h_1$ $P(h_1)$ in the
form
\begin{equation}
\chi = \lim_{h \to 0+} \left \langle \frac{\Delta z}{\Delta h} \right
\rangle \simeq \left \langle \frac{\Delta z_1}{L_z} \right \rangle
\lim_{h \to 0+} P(h_1),
\label{Eqsuskis2}
\end{equation}
because the magnetization of a system $m(h) \simeq \overline{z(h)}/
L_z$, and since the distance in the jump between the minima $\langle
\Delta z_1 \rangle \sim L_z$~\cite{Seppala00}, independently of the
sample-dependent $h_1$. It is expected that a scaling form $P(h_1)
\simeq \langle h_1 \rangle \bar{P}(h_1 / \langle h_1 \rangle )$
applies, and that $P$ remains finite in the limit $h_1 \rightarrow 0$.
Next we compare the average susceptibility as a function of the number
of valleys $N_z$ to the conjecture that in the presence of the field
the average gap for the original and excited state follows an extremal
statistics form similar to Eq.~(\ref{gap}).

\begin{figure}[f]
\centerline{\epsfig{file=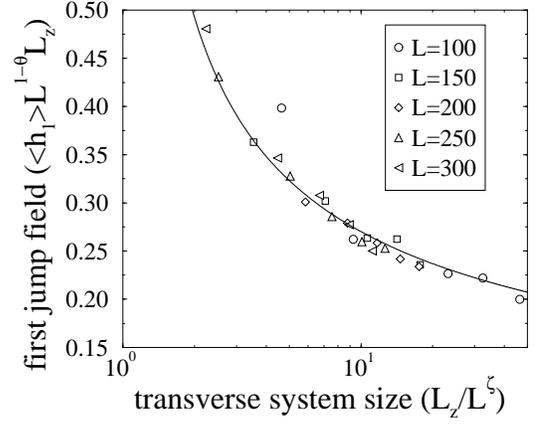,width=6cm,angle=-90}}
\caption{The scaling function $f(y)$ of the scaled disorder-average of
the jump field $\langle h_1 \rangle L^{1-\theta} L_z$ as a function of
scaled transverse system size $L_z/L^{\zeta}$ for the system sizes
$L=$100, 150, 200, 250 and $300$, each with $\bar{z}_0/L_z \simeq
\rm{const}$. $\theta =1/3$, $\zeta =2/3$.  The line has a shape $f(y)
= 0.41 \ln(y)^{-1/2}$. 	Here the number of realizations ranges from
$N=500$ for $L=300$, $L_z=500$ to $N=2600$ for $L=200$, $L_z=600$.}
\label{fig4}
\end{figure}

The simulations are done again using a fixed height window in which
the original ground state without the field is found. After that the
external field $h$ is slowly applied by increasing the coupling
constant values $J_\perp(z) = J_{random} + hz$, where $J_\perp$ is
perpendicular to the $z$-direction, until the first jump is observed
with the corresponding $h_1$ and $\Delta z_1$. In order to find the
scaling relation for the first jump field $h_1$, we make the Ansatz
$\langle \Delta E_1 \rangle = \langle h_1 \rangle L L_z$, since the
field contributes to the polymer energy proportional to $L^D$ ($D=1$)
and $L_z \sim \langle \Delta z_1 \rangle$ is the difference in the
field contributions $hz$ to the energy at finite $h$ at different
average valley heights $z_0$, $z_1$.  Hence
\begin{equation}
\langle h_1(L,L_z) \rangle L L_z \sim L^{\theta} 
f\left(\frac{L_z}{L^{\zeta}}\right),
\label{scalingh1} 
\end{equation}
where the scaling function $f(y)=
[\ln(L_z/L^{\zeta})]^{(\eta-1)/\eta}$.  Fig.~\ref{fig4} shows the
scaling function (13) with a collapse of $\langle h_1(L,L_z) \rangle
L^{1-\theta}L_z$ versus $L_z/L^{\zeta}$ for various $L$ and $L_z$
which is again in good agreement the logarithmic extreme scaling
correction.  Generalizing to arbitrary dimensions one has the behavior
of $\langle h_1(L,L_z) \rangle \sim L^{\theta-D} L_z^{-1} [\ln(L_z/
L^{\zeta}) ]^{-(\eta-1)/\eta}$.  For the susceptibility,
Eq.~(\ref{Eqsuskis2}), one obtains, using $\langle h_1 \rangle$ for
the normalization factor at $P(h_1 = 0)$,
\begin{equation}
\chi \sim L^{D-\theta} L_z [\ln(L_z/L^{\zeta})]^{(\eta-1)/\eta},
\label{Eqsuskis3}
\end{equation}
and in the isotropic limit, $L \propto L_z$, the total susceptibility 
$\chi_{tot} = L^d \chi$  becomes (when $\eta=2$)
\begin{equation}
\chi_{tot} \sim L^{2D+1-\theta} [(1-\zeta)\ln(L)]^{1/2}.
\label{Eqsuskistot}
\end{equation}
Notice that for most random manifolds $1 - \zeta > 0$ with the
exception of 2D random field Ising DW for which $\zeta
\simeq 1$ at large scales~\cite{Seppala98}, and thus the
susceptibility does not diverge~\cite{Aizenman} as the premise $N_z
> 1$ does not hold in that case.  If the condition $N_z > 1$ is
violated the extreme statistics correction disappears.  Thus the
extremal statistics of energy landscapes leads to a logarithmic
multiplier in the susceptibility, Eq.~(\ref{Eqsuskistot}), of the
DW. This result differs from algebraic forms of scaling
\cite{Seppala00}, see also \cite{Shapir}.

To conclude, we have considered the average energy differences or
``gaps'' in the energy landscape of (two-dimensional) elastic
manifolds. An extremal statistics argument in the system geometry with
many independent valleys shows that the ground state energy and the gap
have logarithmic scaling functions, also reproduced with numerical
studies. An illuminating connection can be made to Kardar-Parisi-Zhang
nonequilibrium surface growth.  Finally, we demonstrate that the gap
scaling shows up in the susceptibility of random manifolds. This might
have implications for flux line lattices in high-temperature
superconductors, where a similar problem related to barriers has been
analyzed with the aid of extremal statistics \cite{vinokur}.

This work has been supported by the Academy of Finland's Centre of
Excellence Programme (ETS and MJA). PMD thanks the DOE under contract
DOE-FG02-090-ER45418 for support.



\end{multicols}
\widetext

\end{document}